\documentclass[english,a4paper,prfluids,reprint,amsmath,amssymb,superscriptaddress]{revtex4-2}

\usepackage{graphicx}
\usepackage{bm}

\usepackage[utf8]{inputenc}
\usepackage[T1]{fontenc}
\usepackage{mathptmx}
\usepackage{etoolbox}

\usepackage{hyperref}

\usepackage{siunitx}
\usepackage{xcolor}

\renewcommand{\vec}[1]{\bm{\mathbf{#1}}}

\usepackage{upgreek}

\newcommand{\St}{\mathrm{St}}
\newcommand{\StMED}{\mathrm{St}_{\mathrm{MED}}}
\newcommand{\Reynolds}{\mathrm{Re}}

\newcommand{\tCOLL}{t_{COLL}}

\newcommand{\latin}[1]{{\itshape #1}}

\newcommand{\etal}{\latin{et al.}}

\newcommand{\Eqref}[1]{Eq.~\eqref{#1}}

\newcommand{\Figref}[1]{Fig.~\ref{#1}}

\newcommand\edit[1]{\textcolor{black}{#1}}

\begin{document}

\title{Critical scaling law for the deposition efficiency of inertia-driven particle collisions with a cylinder in high Reynolds number air flow}
\date{\today}

\author{Matthew R. Turner}
\email{m.turner@surrey.ac.uk}
\homepage{http://personal.maths.surrey.ac.uk/st/M.Turner/}
\affiliation{School of Mathematics and Physics, University of Surrey, Guildford, GU2 7XH, United Kingdom}

\author{Richard P. Sear}
\email{r.sear@surrey.ac.uk}
 \homepage{https://richardsear.me/}
\affiliation{School of Mathematics and Physics, University of Surrey, Guildford, GU2 7XH, United Kingdom}

\begin{abstract}
The Earth's atmosphere is an aerosol, it contains suspended particles. When air flows over an obstacle such as an aircraft wing or tree branch, these particles may not follow the same paths as the air flowing around the obstacle. Instead the particles in the air may deviate from the path of the air and so collide with the surface of the obstacle. It is known that particle inertia can drive this deposition, and that there is a critical value of this inertia, below which no point particles deposit. Particle inertia is measured by the Stokes number, St.  We show that near the critical value of the Stokes number, St$_c$, the amount of deposition has the unusual scaling law of exp(-1/(St-St$_c$)$^{1/2}$). The scaling is controlled by the stagnation point of the flow. This scaling is determined by the time for the particle to reach the surface of the cylinder varying as 1/(St-St$_c$)$^{1/2}$, together with the distance away from the stagnation point (perpendicular to the flow direction) increasing exponentially with time. The scaling law applies to inviscid flow, a model for flow at high Reynolds numbers. The unusual scaling means that the amount of particles deposited increases only very slowly above the critical Stokes number. This has consequences for applications ranging from rime formation and fog harvesting to pollination.
\end{abstract}

\maketitle

\section{Introduction}

\begin{figure}
  \centering
 \includegraphics[width=8.2cm]{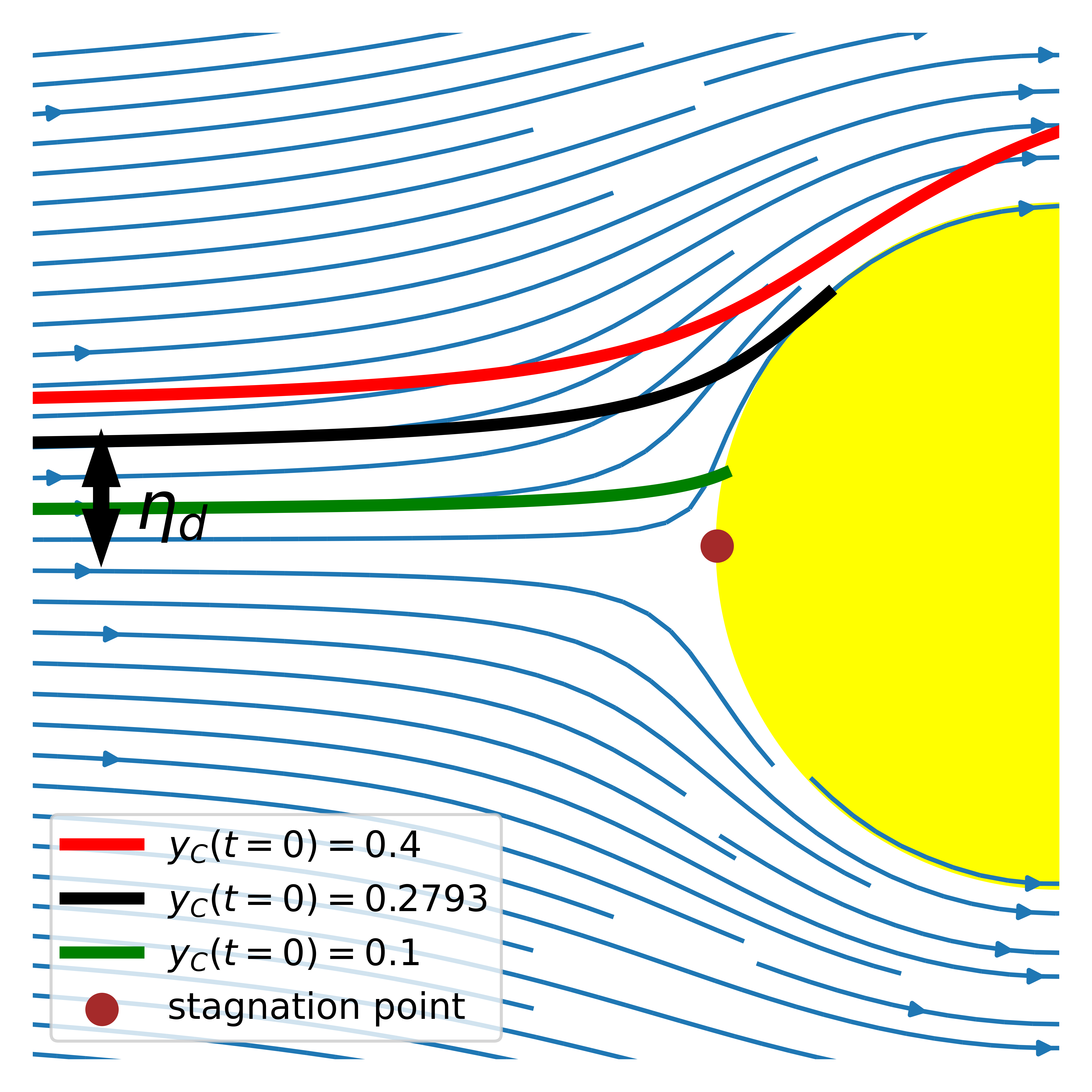}
  \caption{Plot of the cross-section of the cylinder (yellow), the flow field (streamlines in blue), and three trajectories. The green and black trajectories collide with the cylinder, while the red trajectory misses. Here $\St=0.7$ and the initial conditions are at Cartesian coordinate $x_C(t=0)=-10$, with initial $y_C$ values given in the key. The black trajectory is the one that defines the edge of the region where particles collide, and so its value of $y_C=0.2793= \eta_d$. N.B. Numerical errors mean we cannot determine $\eta_d$ to four significant figures, so the $y_C$ value for black curve should not be taken as accurate to four figures. 
  }
  \label{fig:cylinder_collision}
\end{figure}

The Earth's atmosphere is an aerosol, in that it contains suspended particles, with sizes up to tens of micrometres \cite{pruppacher_book}. For example, clouds and fogs are aerosols of water droplets suspended in air. When air flows over an obstacle such as an aircraft wing or tree branch, the suspended particles may follow the air flow around the obstacle, or they may deposit on the surface of the obstacle. For particles tens of micrometres in diameter, deposition on the surface of the obstacle is largely due to the particle's inertia. While the air flow curves to move around the obstacle, the particle's inertia means the particle tends to move in straight lines, and so collides with the obstacle, see \Figref{fig:cylinder_collision}. Here we study these inertia-driven collisions.

Deposition of particles from flowing air onto obstacles occurs in many contexts. These include filtration \cite{wang2013,robinson2021,robinson2022}, harvesting water from fog \cite{parker2001,shahrokhian2020,azeem2020}, pollination \cite{niklas1985,pawu1989}, and rime (ice) formation \cite{makkonen1984,makkonen2000,gao2021}. It has been studied for approximately a hundred years. A lot of the earliest work considered applications where the air was flowing rapidly, at large Reynolds numbers, which is relevant to applications such as rime forming on the leading edges of aircraft wings. The large Reynolds number limit is also the limit we consider here. Rime forms by water droplets below the freezing temperature ($T=0$$~^\circ$C) depositing on a surface and then freezing. This can coat a wing with ice, risking a crash.

In 1931 Albrecht \cite{albrecht1931} found a threshold in the inertia, below which no (point) particles deposited on the obstacle. A minimum amount of inertia is needed before any particles are deposited. Then in the 1940s, first G.I.~Taylor \cite{taylor1940} (also in the scientific papers of G.I.~Taylor\cite{taylor_vol3}), and then Langmuir and Blodgett \cite{langmuir1946} calculated this critical value of the inertia. Here we confirm this result, and build on it. We determine the critical scaling of the deposition efficiency, and show how this relates to the flow field around the obstacle. 

There has been considerable work on this problem since the 1940s \cite{finstad1988comp,finstad1988median}, motivated by its many applications, but the critical scaling has not been studied before, for the high Reynolds number flow field considered by Taylor \cite{taylor1940,taylor_vol3}, and by Langmuir and Blodgett \cite{langmuir1946}. How the threshold varies with Reynolds number has been studied by Phillips and Kaye\cite{phillips1999}, and Ara\'{u}jo~\etal\cite{araujo2006} determined the critical scaling for zero-Reynolds-number flow. 

\edit{
Most of the work has been computational or theoretical, as experiments on collisions in high- Reynolds-number flows are challenging. However,
Wong and coworkers did obtain some experimental data on deposition efficiency at Reynolds numbers of hundreds \cite{wong1955}. This was for an aerosol with particles with a narrow distribution of sizes. They found no measurable deposition below a threshold near that predicted by Taylor \cite{taylor1940,taylor_vol3}, and by Langmuir and Blodgett \cite{langmuir1946}. So these experiments agree with theoretical predictions (for a simplified model) that a threshold exists. Makkonen and coworkers \cite{makkonen1987,makkonen1992,makkonen2018} have measured ice deposition on cylinders. This is for the typical case in the environment, where the droplets have a broad range of sizes, which complicates comparison with theoretical predictions. Here we mostly consider aerosols of identical particles but we do look at the affect of a distribution of particle sizes, in order to compare with this work.
}

\subsection{Obstacles in high Reynolds number flow}

The air flow over an aircraft wing is fast, speeds $U$ of order $\SI{100}{\metre\per\second}$ ($\simeq \SI{360}{\kilo\metre\per\hour}$). This speed, combined with a wing leading edge radius $R$ of order $\SI{10}{\centi\metre}$, means that the Reynolds number of the flow over the wing is
$\Reynolds=UR / \nu\sim 10^6 \gg 1$ with $\nu\simeq 10^{-5} \SI{}{\metre\per\second\squared}$, the kinematic viscosity of air. 
So, following Langmuir and Blodgett and many others, we approximate the airflow over a wing by inviscid, incompressible flow over an infinite cylinder, where we have a simple analytic expression for the flow field \cite{acheson_book}. This is a simple model of high-Reynolds-number flow.

The cylinder is of radius $R$, with its axis along the $z_C$-axis, directed into the page in \Figref{fig:cylinder_collision}. The cylinder is taken to be at rest in the frame of reference. Far from the cylinder, flow is in the $x_C$ direction: $U{\bf i}$. Note that we denote the Cartesian coordinates as $(x_C,y_C,z_C)$, with corresponding unit vectors ${\bf i},~{\bf j},~{\bf k}$, because below we will use $x$ and $y$ to indicate distances from the stagnation point. 
In cylindrical polar coordinates, the flow field $\vec{u}(\vec{r})$ is given  by
\begin{eqnarray}
\frac{\vec{u}}{U}=\left(1-\frac{R^2}{r^2}\right)\cos(\theta)\widehat{\vec{r}}
-\left(1+\frac{R^2}{r^2}\right)\sin(\theta)\widehat{\vec{\theta}},
\label{eq:flowfield}
\end{eqnarray}
where $(r,\theta)$ are plane polar coordinates in the $(x_C,y_C)$-plane. Streamlines of the flow field are illustrated in \Figref{fig:cylinder_collision}.

In the remainder of this paper, we will set the radius of the cylinder $R=1$ and the flow field speed $U=1$. \edit{So for example, both the Cartesian coordinate $x_C$ and the distance along the Cartesian coordinate from the stagnation point, $x$, are in units such that the cylinder radius $R=1$.} This choice also means that time is measured in units such that it takes unit time to move a distance of the cylinder's radius, when moving at speed $U$.

\subsection{Particles in air flowing around an obstacle: The effect of inertia on deposition}

We approximate the particles by point particles,
they are about ten thousand times smaller than the wing leading edge. Point particles that follow the streamlines of fluid flow perfectly never collide with the obstacle. However, if the particles have inertia then when the air flow changes direction to flow around the obstacle, the particle's inertia may cause it to go straight on potentially leading it to crash into the obstacle. So particle inertia can cause collisions.

The inertia of a particle is quantified by its Stokes number $\St$, defined for a particle of mass $m_p$ by
\begin{equation}\label{eq:stokes-number}
  \St = m_p U B \big/  R
\end{equation}
with $B$ the particle mobility and $R$ the lengthscale of the obstacle.
The Stokes expression for the mobility is $B = 1/(6\pi \eta a_p)$ with $\eta$ the dynamic viscosity of air, and $a_p$ the radius of the particle.
The Stokes number is the dimensionless ratio between the inertia of a particle -- which tends to cause the particle to move in straight lines -- and the friction between the particle and the surrounding air -- which tends to make the particle follow streamlines.
For a wing of width $\SI{0.1}{\metre}$, the Stokes number is of order one for droplets micrometres in diameter, so cloud droplets of size micrometres, and tens of micrometres will deposit on the wing.

In the $\St\to\infty$ limit, particles move in straight lines and so in that limit a cylinder of radius $R$ sweeps out a strip of air of thickness $2R$, collecting all the particles in this strip.
This allows us to define the deposition or collection efficiency
$$
\eta_d=\left. \begin{array}{l}
     \mbox{maximum displacement from cylinder axis}\\
     \mbox{perpendicular to the flow direction, for}\\
     \mbox{which a particle collides with cylinder surface}  
\end{array}\middle/ R \right.   
$$
which varies from zero when no particles collide, to one when
$\St\to\infty$. The displacement is taken far upstream of the cylinder. Calculation of $\eta_d$ is done by starting particle trajectories far upstream of the cylinder at varying values of the displacement $y_C$ normal to the flow direction. Then $\eta_d$ is defined by the largest initial displacement along $y_C$, for which the particle collides. This is illustrated by the black trajectory in \Figref{fig:cylinder_collision}.

\begin{figure}
  \centering
 \includegraphics[width=\linewidth]{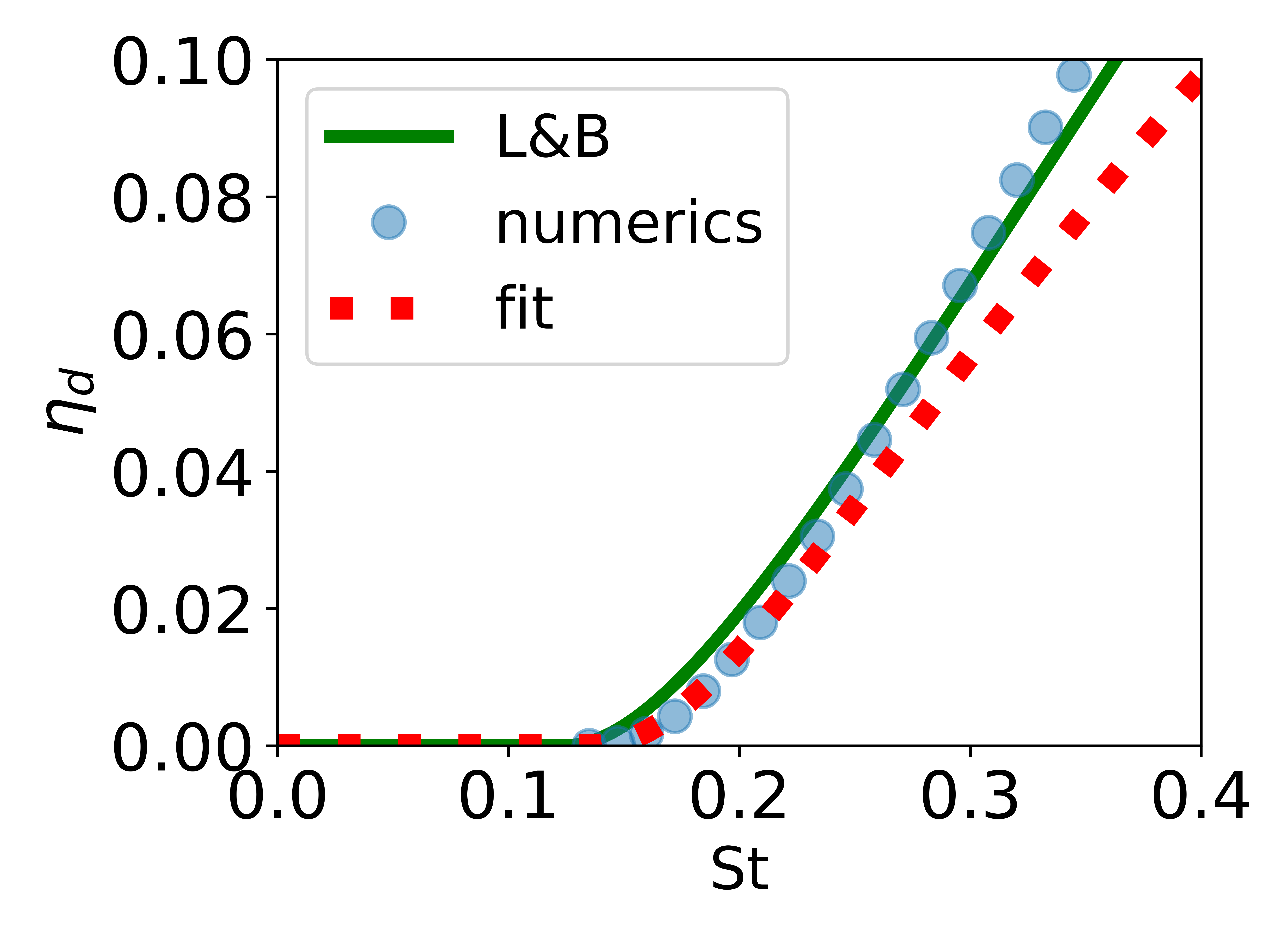}
  \caption{Plot of the deposition or collection efficiency $\eta_d$ as a function of the Stokes number. Shown are numerical results (blue circles), the function of Langmuir and Blodgett \cite{langmuir1946} (green curve) and our fit to the region of small $\delta=\St-\St_c$ (dashed red curve). 
  }
  \label{fig:width_vs_St}
\end{figure}

Numerical results for $\eta_d$ as a function of $\St$, together with the fit of Langmuir and Blodgett \cite{langmuir1946}, are shown in \Figref{fig:width_vs_St}. Note that at small values of St, the collection efficiency increases very slowly with increasing inertia. This is what we will explain here. It follows directly from the high Reynolds number flow field. In the low-Reynolds-number limit, where the flow field is very different, $\eta_d$ increases much more rapidly above the critical Stokes number \cite{araujo2006,robinson2021,robinson2022b}. Langmuir and Blodgett's fit is in Appendix \ref{sec:appLB}, and the details of our numerical calculations are in Appendix \ref{sec:appnumerics}. See Appendix \ref{sec:appscaling} for the details of the fit in \Figref{fig:width_vs_St}.


\section{Newton's equation for a particle in flowing air}

For a particle suspended in flowing air we assume that the only force on the particle is the friction with the surrounding air, which is taken to be proportional to the difference between particle's velocity $\vec{v}$ and the local flow velocity $\vec{u}$. This force is taken to act on the particle's centre of mass. Then
Newton's equation for the particle motion 
is
\begin{equation}\label{eq:stokes-newton}
  \St \frac{d \vec{v}}{d t} = - (\vec{v} - \vec{u}).
\end{equation}
In cylindrical polar coordinates, this is
\begin{subequations}\label{eq:stokes-newton2}
  \begin{align}
  \St \left(\ddot{r}-r\left(\dot{\theta}\right)^2\right)
  =-\left(\dot{r}-u_r\right),
    \\
  \St \left(\ddot{\theta}+\frac{2\dot{r}\dot{\theta}}{r}\right)
  =-\left(\dot{\theta}-u_{\theta}\right),
  \end{align}
 \end{subequations}
where $\vec{u}=u_r\hat{\vec{r}}+u_\theta\hat{\vec{\theta}}$.

\section{The cylinder's forward stagnation point}

We are interested in the behaviour near the critical Stokes number of Langmuir and Blodgett \cite{langmuir1946}. Here, particle trajectories pass close to the stagnation point at the front of the cylinder. This stagnation point is at $r=1$ and $\theta=\pi$, see \Figref{fig:cylinder_collision}. So we will study behaviour near this stagnation point. We start by changing variables to the distance to contact with the cylinder $x = r - 1\ll1$, and the angle from the angle of the stagnation point $y = \pi-\theta\ll1$. Note that $x$ and $y$ are not the conventional Cartesian coordinates, which we denote by $x_C$ and $y_C$.

In these new coordinates, Newton's equations for the particle, \Eqref{eq:stokes-newton2}, become
\begin{subequations}
  \begin{align}
    \ddot{x} - (1 + x) \dot{y}^2
    &=
    - \frac{\dot{x} - u_r}{\St},
    \\
    \ddot{y} + \frac{2}{1 + x} \dot{x} \dot{y} 
    &=
    - \frac{\dot{y} + u_\theta}{\St}.
  \end{align}
  \label{eq:xydot}
\end{subequations}
Note that $\dot{y}=-\dot{\theta}$ and
$\ddot{y}=-\ddot{\theta}$ etc. 
Near the stagnation point, we can expand the flow field in \Eqref{eq:flowfield} as a series in $x$ and $y$
\begin{eqnarray}
\vec{u}&=&\left(-2x + 3x^2+ \mathcal{O}({\rm cubic~terms})\right)\hat{\vec{r}}\nonumber\\
&&+\left(-2y +2xy+ \mathcal{O}({\rm cubic~terms})\right)\hat{\vec{\theta}}.
\label{eq:flowfieldxy}
\end{eqnarray}
It is worth noting that as we do not have stick boundary conditions here, $u_{\theta}$ is not zero at the cylinder surface, except at the stagnation point; $u_{r}$ is zero at the surface is because there is no flow into the cylinder.

We substitute the flow field of \Eqref{eq:flowfieldxy} into the equation for the particle trajectory (\Eqref{eq:xydot}). Then if we keep only linear and quadratic terms we obtain
\begin{subequations}
  \begin{align}
    \ddot{x} - \dot{y}^2 &=
    - \frac{1}{\St} \left( \dot{x} + 2x -3x^2\right),
    \label{eq:odex_exp}
    \\
    \ddot{y} + 2 \dot{x} \dot{y} &=
    - \frac{1}{\St} \left( \dot{y} - 2y + 2xy\right),
    \label{eqn:y}
  \end{align}
\end{subequations}
\edit{We now solve these equations to find out which are the trajectories of particles that collide with the cylinder.}

\section{Particle trajectories on axis, and the critical Stokes number}

We start by considering particle trajectories precisely on axis, where $\theta=\pi$ and $y=0$. This will enable us to determine the critical value of the Stokes number, below which no particles collide with the cylinder. We follow Taylor \cite{taylor1940,taylor_vol3}, and Langmuir and Blodgett \cite{langmuir1946} here. On axis the system reduces to a one dimensional problem. 

Near the stagnation point $x\ll 1$, and we can approximate the flow by keeping only the leading order terms in $x$. Then \Eqref{eq:odex_exp} is just
\begin{equation}
     \St\ddot{x}
    =
    -\dot{x} - 2x,
    \label{eqn:xdd}
\end{equation}
which as both Taylor, and Langmuir and Blodgett realised,  is just the differential equation for damped simple harmonic motion (SHM). It has solutions
\begin{equation}
    x(t)=A_0\exp\left(\lambda_1t\right)+B_0\exp\left(\lambda_2t\right),
    \label{eq:shm1}
\end{equation}
\begin{equation}
    \lambda_1,\lambda_2=\frac{-1\pm\sqrt{1-8\St}}{2\St},
    \label{eq:lambdas}
\end{equation}
where $A_0$ and $B_0$ are fixed by the initial conditions, for example $x(t=0)=x_0$ and $\dot{x}(t=0)=u_0$. See Appendix \ref{sec:app1dshm}.

For $0<\St<1/8$ both $\lambda$ are real and negative (overdamped SHM solutions), and so the particle approaches the cylinder surface at a speed which decays exponentially. There is only a collision in the $t\to\infty$ limit, i.e., no collision at finite time.  But if $\St>1/8$ then we have complex $\lambda$ (underdamped SHM solutions), and the collision will occur in finite time. The critical value of the Stokes number is therefore $\St_c=1/8$. We can define the distance from this critical value as
\begin{equation}
    \delta=\St-\St_c.
    \label{eqn:delta}
\end{equation}
Note that there are initial conditions for \Eqref{eq:shm1}, for which the particle collides in finite time\cite{ingham1990}. However, they do not appear to be physically relevant,  as Ingham \etal\cite{ingham1990} discuss.

\subsection{Time to collision, on axis}

The collision occurs when $x(\tCOLL)=0$
and is set by the angular frequency --- the imaginary part of \Eqref{eq:lambdas} ---  $\omega=\sqrt{1-8\St}/(2\St)\simeq 8\sqrt{2}\delta^{1/2}$. As expected
(see Appendix \ref{sec:app1dshm} for details) the time to collide $\tCOLL$ is half the period
\begin{equation}
    \tCOLL\simeq \frac{\pi}{8\sqrt{2}}\delta^{-1/2}~~~~~~\delta\ll 1.
    \label{eqn:t_collide1D}
\end{equation}
As the critical Stokes number is approached from above, the time to reach the cylinder surface and collide diverges as $1/\delta^{1/2}$. Numerical calculations using the full flow field agree with this observation, see Appendix \ref{sec:appscaling}.

\begin{figure}
  \centering
 \includegraphics[width=8.2cm]{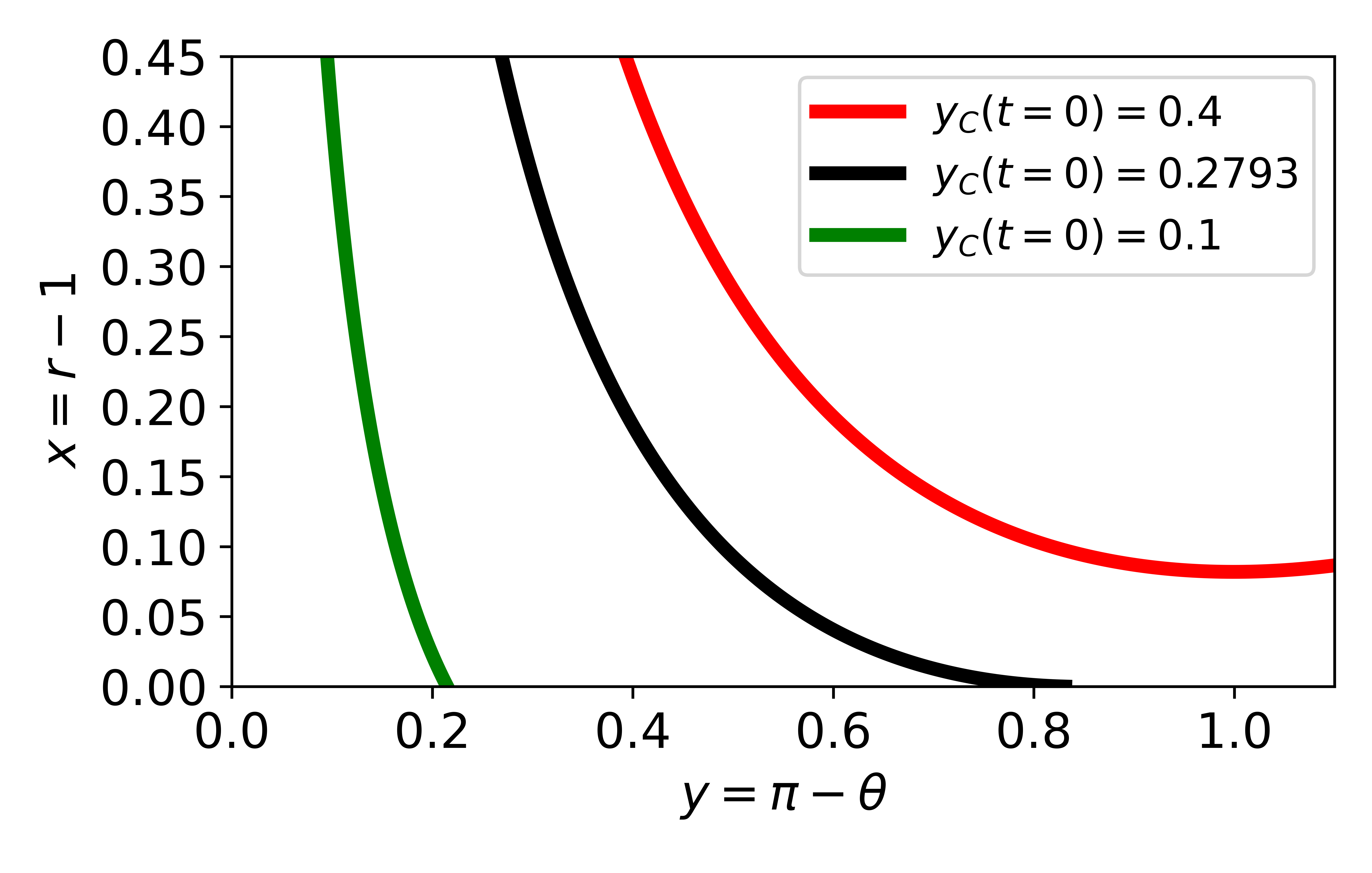}
  \caption{Plot of the trajectories in \Figref{fig:cylinder_collision} but in the $xy$ plane. The green and black trajectories collide with the cylinder, while the red trajectory misses. Here $\St=0.7$ and the initial conditions are at Cartesian coordinate $x_C(t=0)=-10$, with initial $y_C$ values given in the key. Note that the trajectory which just collides does so tangentially to the surface.
  }
  \label{fig:cylinder_collision_xy}
\end{figure}

\section{Particle trajectories off axis, and the critical scaling of the deposition efficiency}

We now consider the full two-dimensional case, near the stagnation point. Off axis we also need an equation for $y$. Retaining only the leading order linear terms in \Eqref{eqn:y} for $y$ leads to
\begin{equation}
        \St\ddot{y} =
    - \dot{y} + 2y,
    \label{eqn:ydd}
\end{equation}
in which there is no coupling with the $x$ direction. This equation is almost the damped SHM equation again, but in SHM language, the \lq force\rq~term has the opposite sign, so it is not a restoring force but drives exponential growth of $y$. The solution is
\begin{equation}
    y(t)=C_0\exp\left(\mu_1t\right)+D_0\exp\left(\mu_2t\right)
        \label{eq:yexp}
\end{equation}
with
\begin{equation}
    \mu_1,\mu_2=\frac{-1\pm\sqrt{1+8\St}}{2\St}
\end{equation}
Now, $\mu_1>0$ and $\mu_2<0$ so at long times the $\mu_1$ solution dominates; it increases exponentially while the other solution decays to zero. The constants $C_0$ and $D_0$ are related to the initial conditions. If $y(t=0)=y_0$ and $\dot{y}(t=0)=v_0$, then
\begin{equation}
    C_0=\frac{v_0-\mu_2y_0}{\mu_1-\mu_2}~~~{\rm and}~~~D_0=-\frac{v_0-\mu_1y_0}{\mu_1-\mu_2}.
\end{equation}



Having determined the leading order behaviour of $y$, we return to \Eqref{eq:odex_exp} for $x$.
The leading order $y$ term in \Eqref{eq:odex_exp} is the $\dot{y}^2$ term. Retaining only this term, we have
\begin{equation}
    \ddot{x}    + \frac{\dot{x} + 2x}{\St} =\dot{y}^2.
    \label{eq:xodeminimal}
\end{equation}
The new term acts like an effective force that pushes the particle away from a collision.

The general solution to \Eqref{eq:xodeminimal} is given in \Eqref{eqn:gensol2} in Appendix \ref{appen:gsol}. We take this equation, expand for $\delta\ll1$, keep only leading-order terms, and then set $x=0$ at the collision time of  \Eqref{eqn:t_collide1D}. This yields
\begin{eqnarray}
    x(\tCOLL)&=&-A_0\exp\left[-\frac{\pi}{2\sqrt{2}}\delta^{-1/2}\right]
    \nonumber\\
    &&+E_0\exp\left[\frac{(\sqrt{2}-1)\pi}{\sqrt{2}}\delta^{-1/2}\right]=0,
    \label{eqn:x3}
\end{eqnarray}
with $E_0=C_0^2(11-6\sqrt{2})/49$. This equation is a function of $\delta$ and the initial boundary conditions on $x$ and $y$.

Equation (\ref{eqn:x3}) has two terms. The first term is negative and is the same for the on-axis case. It is this term that results in a collision after a time $\sim 1/\delta^{1/2}$. The second term is new for the off-axis case, it {\em increases} exponentially with $\delta^{-1/2}$, and the prefactor scales with $C_0^2\sim (v_0-\mu_2y_0)^2$, i.e., with the initial conditions on $y$.

We want to estimate $\eta_d$, which is set by the largest displacement $y_0$ for which the particle collides with the cylinder, which is when \Eqref{eqn:x3} is satisfied. Near the critical Stokes number $y_0$ will be small, and the exponential variation of the terms in \Eqref{eqn:x3} suggests it needs to be exponentially small, so we write $C_0=\zeta\exp[\alpha(\delta)]$ where $\zeta$ is a constant to leading order in $\delta$, and $\alpha$ is a function of $\delta$. 

Then we insert $C_0=\zeta\exp(\alpha)$ into \Eqref{eqn:x3}. For small $\delta$ solutions are only possible when the exponential exponents are equal. Equating the exponents of the two terms gives us
\begin{equation}
    -\frac{\pi}{2\sqrt{2}}\delta^{-1/2}=2\alpha+\frac{(\sqrt{2}-1)\pi}{\sqrt{2}}\delta^{-1/2},
\end{equation}
and
\begin{equation}
    \alpha=-\frac{(4-\sqrt{2})}{8}\pi\delta^{-1/2}\simeq -1.015\delta^{-1/2}.
    \label{eq:alpha}
\end{equation}



Collisions only occur at small $\delta$ when $y_0$ scales as $\exp(-1.015\delta^{-1/2})$. This sets the width of the strip of air over which particles collide with the cylinder, and so the deposition efficiency is
\begin{equation}
   \eta_d\sim\left\{
   \begin{array}{cc}
   0 & \delta \le 0 \\
   \exp\left[-\frac{(4-\sqrt{2})}{8}\pi\delta^{-1/2}\right] & 0<\delta \ll 1
   \end{array}\right.
   \label{eq:etad_scaling}
\end{equation}
Above the critical Stokes number, the collection efficiency has the unusual scaling $\exp(-1/\delta^{1/2})$. This means that the deposition efficiency increases only slowly above the critical Stokes number, see \Figref{fig:width_vs_St}. Numerical calculations with the full flow field confirm the result. They are in Appendix \ref{sec:appscaling}. Limited experimental data \cite{wong1955} near $\St_c$ means that a quantitative comparison with our scaling result is not possible.

The scaling is a direct consequence of the form of the flow field near the stagnation point, \Eqref{eq:flowfieldxy}. From the fact that the flow field varies linearly with $x$ and $y$. The flow field leads to the time to collision scaling as $1/\delta^{1/2}$, and to the $y$ coordinate increasing exponentially with time, which in turn gives us \Eqref{eq:etad_scaling}.

Numerical results for particle trajectories as functions of $x$ and $y$ are plotted in \Figref{fig:cylinder_collision_xy}. The trajectory that just collides, and so defines $\eta_d$, is in black. Note the rapidly increasing $y$ as the collision is approached, and that the collision occurs tangential to the cylinder surface.

\section{Comparison with particle deposition in Stokes flow}

Our results are for a flow field that neglects viscosity, in effect the infinite-Reynolds-number limit. The opposite limit is where inertia is negligible and viscosity dominates. This is the zero-Reynolds-number or Stokes-flow limit, and it is relevant to the filtration of aerosol particles from air by fibrous filters, where the Reynolds number is small \cite{robinson2021,wang2013,riosdeanda2022}.
For particles in Stokes flow, there is also a critical value of the Stokes number, but the scaling of the deposition efficiency is $\eta_d\sim \delta^{1/2}$, which is completely different scaling. Langmuir and Blodgett suspected that there were critical Stokes numbers for a number of different flow fields including spheres in Stokes flow, but this scaling was first found for spheres in Stokes flow by Ara\'{u}jo \etal\cite{araujo2006}.

The particle trajectories are very different for the inviscid and Stokes flow fields. For example, with the slip boundary conditions in the inviscid limit, the trajectories where the particle just collides with the cylinder surface do so tangentially, see \Figref{fig:cylinder_collision_xy}. This is because the particle's velocity normal to the cylinder surface tends to zero at the collision, while the tangential component is non-zero. Particles in Stokes flow do not collide tangentially \edit{because here the stick boundary conditions mean that the tangential fluid-flow velocity is zero at contact
\cite{robinson2022b}.}

\edit{At intermediate Reynolds numbers, there will be a boundary layer of thickness $\sim 1/$Re$^{1/2}$. Within this boundary layer the flow field is approximately viscous flow, outside it is closer to inviscid flow. So as the Reynolds nunber is increased the boundary layer thins and the behaviour changes continuously from the Stokes to the inviscid limit. Work by Robinson and coworkers suggests that the deposition behaviour changes smoothly between the limits \cite{robinson2022b}}.

\begin{figure}
  \centering
 \includegraphics[width=\linewidth]{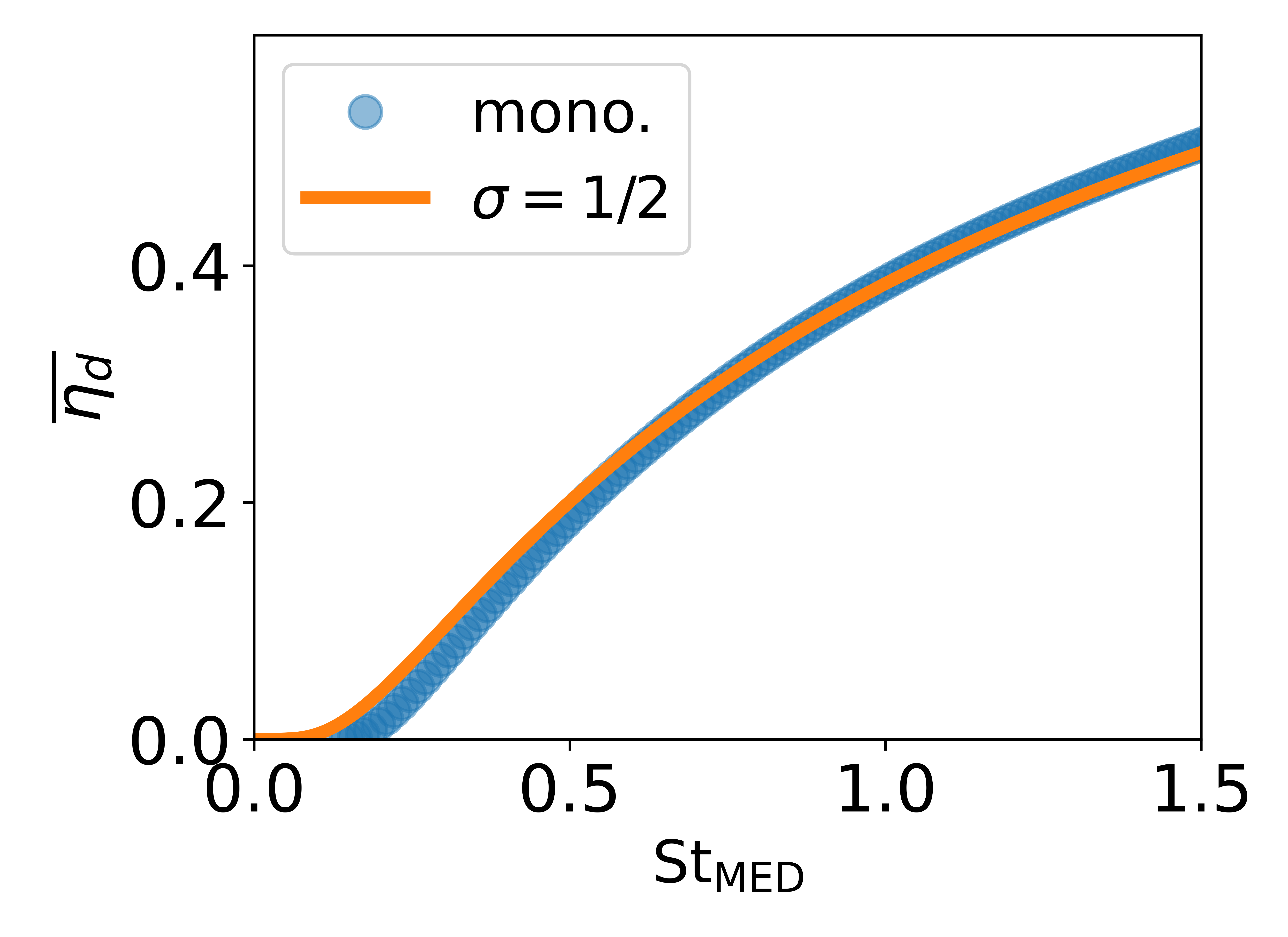}
  \caption{\edit{Plot of the mean deposition or collection efficiency, $\overline{\eta_d}$, as a function of the median Stokes number. Shown are numerical results for monodisperse droplets (blue circles), and for polydisperse droplets whose Stokes numbers obey a log-normal distribution with width parameter $\sigma=1/2$ (orange curve).}
  }
  \label{fig:poly_eta}
\end{figure}

\section{Deposition from an aerosol of particles with a range of diameters}

\edit{In natural aerosols, the particles are almost never all of the same diameter. A broad distribution of particle sizes is typical \cite{makkonen1987,makkonen2018,pruppacher_book}. Within our simplified model, each particle is characterised by a single parameter: the Stokes number. A distribution of particle sizes results in a distribution of Stokes numbers. The Stokes number of a particle is given by \Eqref{eq:stokes-number}. When mass $m_p\propto a_p^3$ and the Stokes mobility $B\propto 1/a_p$, the scaling of the Stokes number is St $\propto a_p^2$, i.e., the Stokes number scales with the square of the particle radius. Thus a distribution of particle radii gives a distribution of Stokes numbers. An aerosol of particles with a range of radii will have particles with a distribution of Stokes numbers: $p(\St)$.}

\edit{We need a model probability distribution function, $p$. We use the standard log-normal distribution}
\begin{equation}
p(\St ; \StMED,\sigma)=\frac{1}{\St\sigma(2\pi)^{1/2}}
\exp\left[-\frac{(\ln\St-\mu)^2}{2\sigma^2}
\right]
\label{eq:log-norm}
\end{equation}
\edit{Here, the median Stokes number $\StMED=\exp(\mu)$ provides a measure of the typical Stokes number. The width of the distribution is set by $\sigma$: the ratio of the standard deviation to the median value is $\exp(\sigma^2/2)(\exp(\sigma^2)-1)^{1/2}$.}

\edit{Within our model, the particles are independent. So the mean deposition efficiency, $\overline{\eta_d}$, for the aerosol is simply}
\begin{equation}
\overline{\eta_d}(\StMED,\sigma)=\int \eta(\St)p(\St;\StMED,\sigma){\rm d St}
\label{eq:poly_eta}
\end{equation}

\edit{In \Figref{fig:poly_eta}, we compare the mean deposition efficiency as a function of median Stokes number, of monodisperse particles, and particles with a log-normal distribution with width parameter $\sigma=1/2$. This distribution has a ratio of standard deviation to the median of $0.60$.}
\edit{Near the critical value of the Stokes number, the convolution of deposition efficiency with the distribution of Stokes numbers of the particles smooths out the transition at $\St_c$. For a distribution of particle sizes and hence and Stokes numbers, the deposition efficiency is never zero, unless {\em all} particles have Stokes numbers less than $\St_c$. Thus when the median Stokes number is below $\St_c$ polydisperse aerosols have larger deposition efficiencies than monodisperse ones. But regardless of polydispersity the deposition efficiency tends to one at larges values of the median Stokes number. So at large median Stokes numbers a broad distribution of particle sizes has little effect on the mean deposition efficiency.}

\begin{figure}
  \centering
 \includegraphics[width=\linewidth]{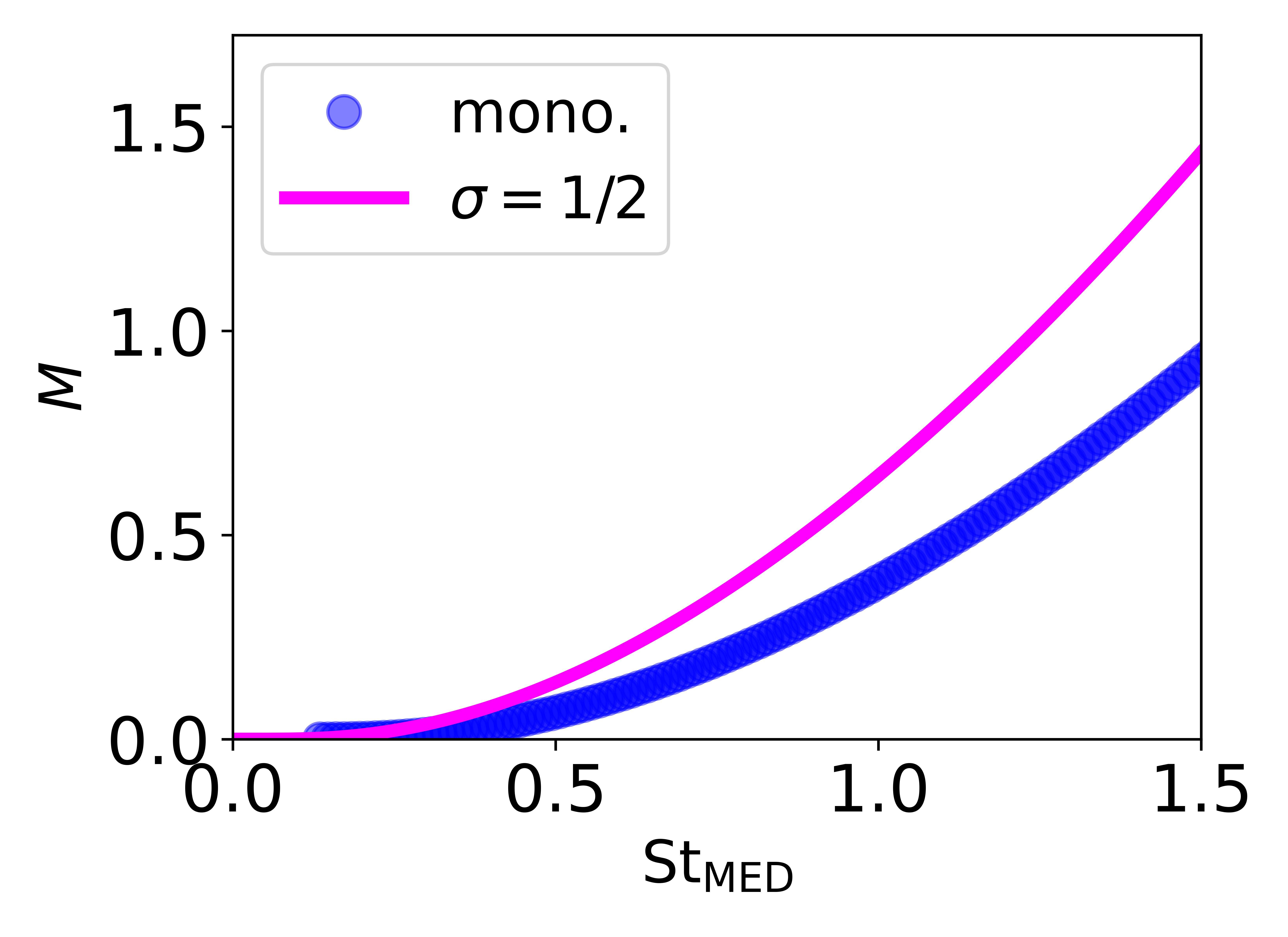}
  \caption{\edit{A quantity $M$, which is proportional to the mass deposited, plotted as a function of the median Stokes number. Shown are numerical results for monodisperse droplets (dark blue circles) and for polydisperse droplets whose Stokes numbers obey a log-normal distribution with width parameter $\sigma=1/2$ (magenta curve).}
  }
  \label{fig:poly_mass}
\end{figure}

\edit{In experiment, the total mass deposited on an obstacle can be measured \cite{makkonen1987,makkonen2018}. As the mass of a particle scales as $a_p^3$, the mass of a particle scales as St$^{3/2}$. So the three-halves moment of the efficiency times the distribution of particle Stokes numbers gives us a quantity proportional to the total mass deposited. We call this quantity $M$:}
\begin{equation}
M(\StMED,\sigma)=\int \eta(\St)p(\St;\StMED,\sigma)\St^{3/2}{\rm d St}
\label{eq:poly_mass}
\end{equation}
\edit{
We have plotted this deposited mass as a function of median Stokes number in \Figref{fig:poly_mass}. As with the deposition efficiency in \Figref{fig:poly_eta}, when a range of particle sizes are present, the transition is smoothed over. Mass is deposited at all values of the median Stokes number. This is consistent with the work of Makkonen and coworkers \cite{makkonen1987,makkonen2018}, who find that some mass is deposited under all their experimental conditions. However, the mass deposited is always larger than for monodisperse droplets with the same median Stokes number. The total mass of aerosol particles is larger in the polydisperse case. (at the same median Stokes number). In addition, the large Stokes number tail of the distribution contributes large amounts to the mass deposited as here the deposition efficiency is high and these large particles contribute large amounts to the total mass deposited.}

\section{Conclusion}

We have built on Taylor's \cite{taylor1940,taylor_vol3}, and Langmuir and Blodgett's work \cite{langmuir1946} to quantitatively understand the behaviour of the deposition efficiency $\eta_d$ near the critical value of the Stokes number. Just above the critical value,  $\eta_d\sim\exp(-1/\delta^{1/2})$. This unusual scaling comes from the fact that for an off-axis particle trajectory, the particle's displacement from the cylinder's axis increases exponentially with time, while the time to reach the cylinder scales as $1/\delta^{1/2}$. Thus unless the initial displacement from the axis is exponentially small, the particle misses the cylinder. The scaling follows directly from the (inviscid) flow field we used; the behaviour of particles in Stokes flow with stick boundary conditions is very different \cite{araujo2006,robinson2021,robinson2022b}. The lesson here is that particle deposition from flowing air is very sensitive to details of the flow field, especially near the stagnation point. The slow increase in deposition efficiency follows from the slip boundary conditions. 
\edit{Particle deposition has varied applications, from rime formation on aircraft wings v to pollination \cite{niklas1985,pawu1989}. The careful analysis of particle trajectories near the forward stagnation point will greatly contribute to the understanding of this issue.}
 
Here, we have studied a cylinder, but Langmuir and Blodgett \cite{langmuir1946} also found a critical value of the Stokes number for spheres in inviscid, axisymmetric flow. This was at the slightly lower value of $\St_c=1/12$. It is straightforward to show that the scaling laws found here for a cylinder are the same (up to multiplicative constants) for a sphere. The details are in Appendix \ref{appen:sphere}.
Our results for the scaling were obtained by expanding around the stagnation point, it is possible that they are general to at least most inviscid flows impinging on a locally parabolic surface in two or three dimensions. 


\section{Supplementary Material}

The supplementary material is a Python Jupyter notebook that performs all numerical calculations, and produces all figures in this work.

\begin{acknowledgements}
It is a pleasure to thank Josh Robinson and Patrick Warren for inspiring and useful discussions. The data that support the findings of this study are available within the article and in a Python Jupyter notebook in supplementary material.
\end{acknowledgements}

\appendix

\section{Langmuir and Blodgett expression for deposition efficiency}
\label{sec:appLB}

The expression used by Langmuir and Blodgett \cite{langmuir1946} is
\begin{equation}
    \eta_d=\left\{
    \begin{array}{cc}
    0 & \St \le 1/8 \\
    0.466\left[\log_{10}(8\St)\right]^2 & 1/8 < \St < 1.1 \\
    \St/(\St+\pi/2) &  1.1 < \St
    \end{array}
    \right.
    \label{eq:LB}
\end{equation}
Note that the functional form near $\St=1/8$ is incorrect but that it is quite close to numerical calculations. This expression, and closely related ones, have been and are used extensively to model ice accretion on the leading edges of surfaces moving through air \cite{makkonen1984,finstad1988comp,finstad1988median,azeem2020}. Finstad \etal\cite{finstad1988median,finstad1988comp} summarise work up to 1998 on improving the approximation of \Eqref{eq:LB}. None of the improvements has the correct functional form for small $\delta=\St-1/8$ but it should be said the work was focused on producing empirical expressions that are accurate over a wide range of values of the Stokes number, not on obtaining the correct value near the dynamical transition.


\section{Details of numerical calculations}
\label{sec:appnumerics}

The numerical calculations were done with a Python Jupyter notebook available in the supplementary material. Particle trajectories were calculated as follows. Particles were started at Cartesian coordinates upstream of the cylinder $(x_{C0},y_{C0})$, i.e., a distance $x_{C0}$ along the $x_C$-axis (the flow direction), and displaced a perpendicular distance $y_{C0}$ off axis. All calculations here are for $x_{C0}=-10$; we checked that increasing the distance upstream further to $-20$ had very little effect. Trajectories were obtained by integrating Newton's equation of motion (\eqref{eq:stokes-newton}), and checking for collisions, which occur whenever $r=1$.

\begin{figure}
  \centering
 \includegraphics[width=\linewidth]{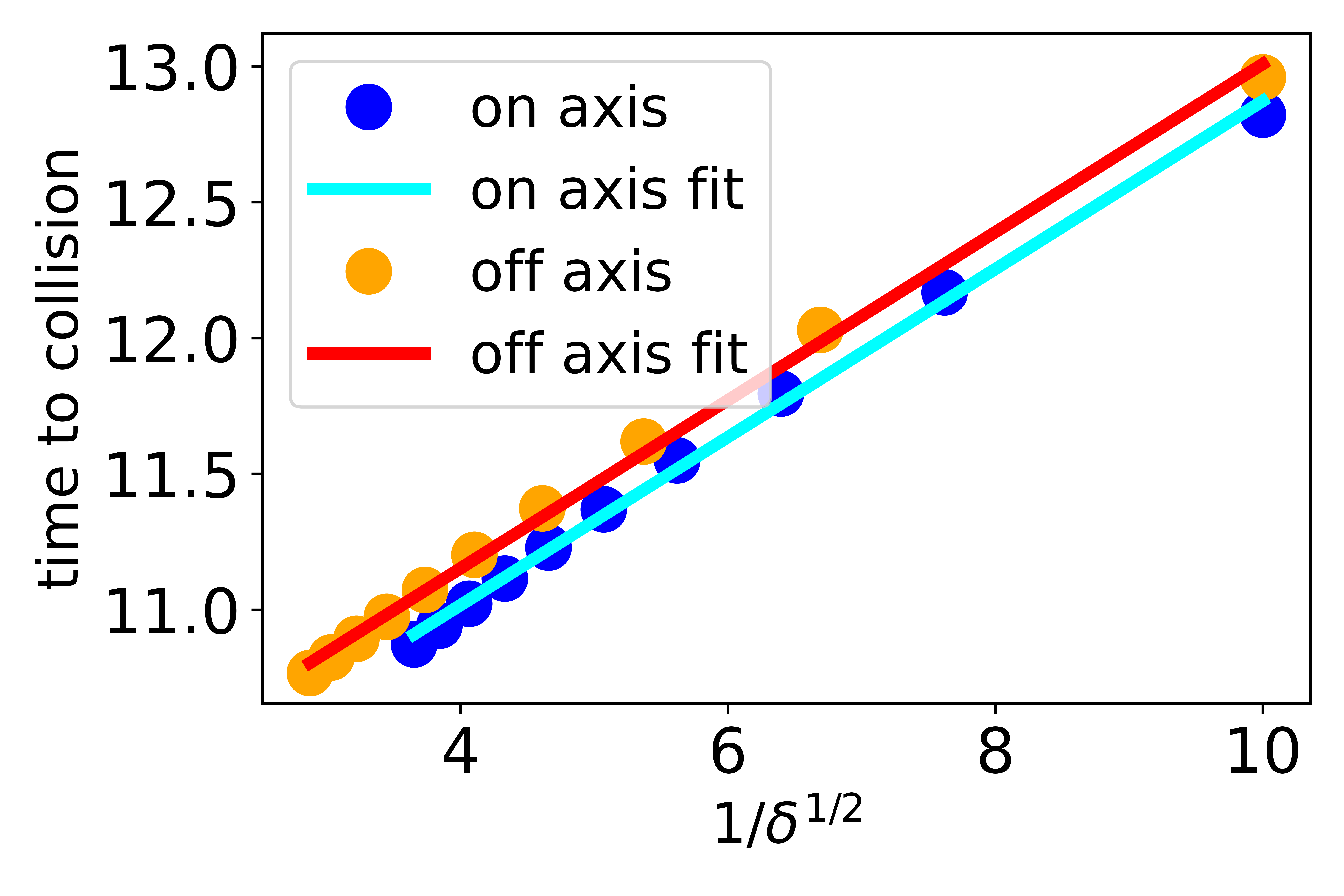}
  \caption{Plot of time to collision, as a function of $\delta=\St-\St_c$. The points are numerical results, and the lines are fits of function $A+B\delta^{-1/2}$. Blue points and cyan line are for on axis collisions, with best fit values $A=9.78$ and $B=0.309$. Orange points and red line are for collisions at the edge of the deposition zone, with $A=9.92$ and $B=0.309$. Note that both values for $B$ are close to our prediction of $t_{COLL}=\pi/(8\sqrt{2}\delta^{1/2})=0.278/\delta^{1/2}$.
  }
  \label{fig:t_collide}
\end{figure}

\subsection{Verification of scaling found by analytical mathematics, by comparison to numerical calculations}
\label{sec:appscaling}

On axis, our analytical results predict that as the critical Stokes number is approached from above, the time to reach the surface of the cylinder and collide, scales as $1/\delta^{1/2}$.
There is another contribution, which is the time for the particle to go from its initial position upstream to a point where $x\ll 1$. That contribution is approximately equal to initial displacement$/U$.

So, we fit a function of the form $A+B\delta^{-1/2}$ to numerical results for the time to collision. The numerical results and fit are shown in \Figref{fig:t_collide}. We see excellent agreement.

We perform the same analysis for the trajectories that define $\eta_d$, at the edge of the strip of air over which collisions occur. The results are in \Figref{fig:t_collide}, and again the agreement is excellent.

\begin{figure}
  \centering
 \includegraphics[width=\linewidth]{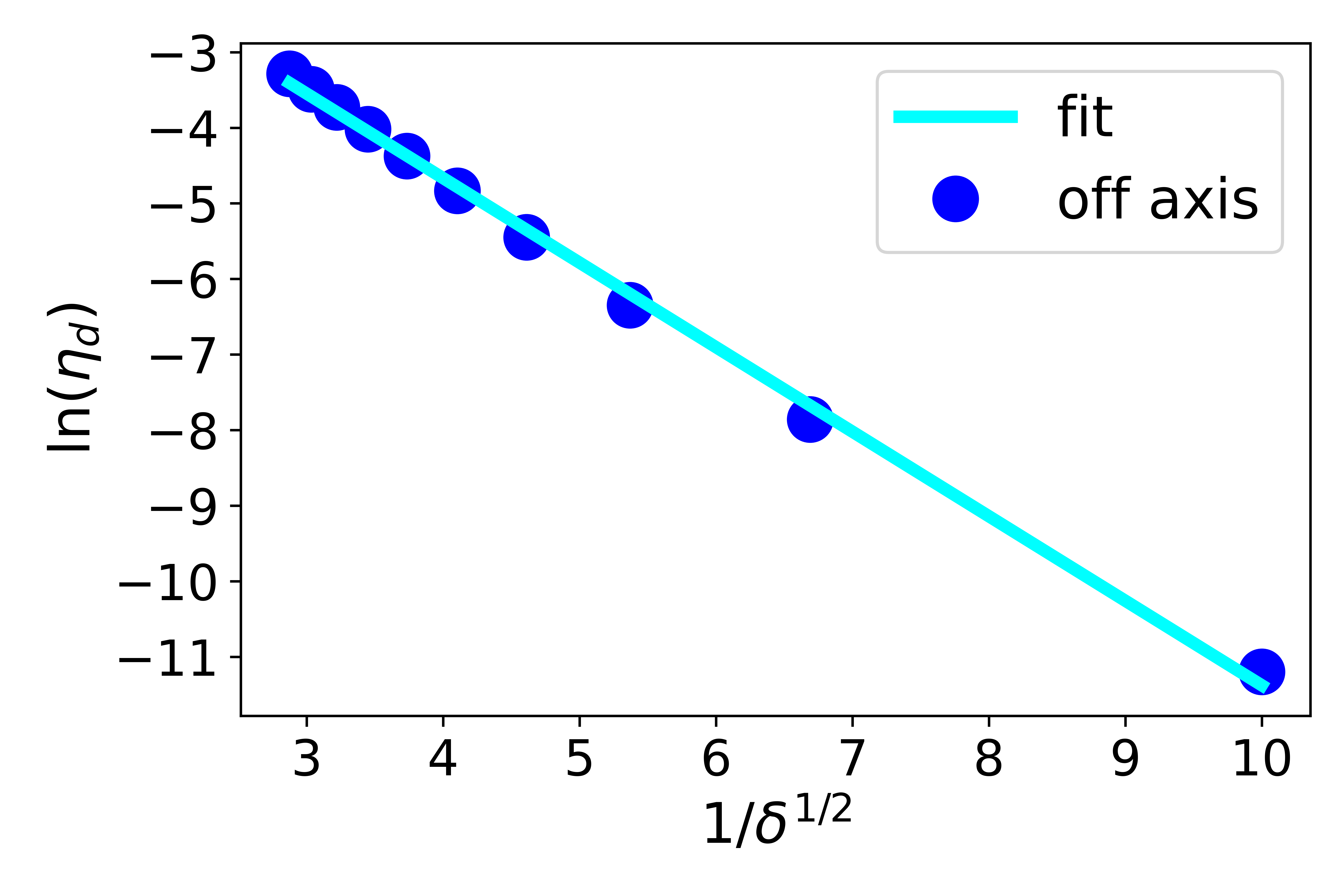}
  \caption{Plot of $\ln(\eta_d)$ as a function of $1/\delta^{1/2}$, to show $\exp(-1/\delta^{1/2})$ scaling of the deposition efficiency. Points are numerical results, line is fit of function $A+B(1/\delta^{1/2})$ with best fit values $A=-0.193$ and $B=-1.118$. Note that value for $B$ is close to our prediction of $\alpha=-1.015/\delta^{1/2}$ (\Eqref{eq:alpha}).
  }
  \label{fig:width_scaling}
\end{figure}

We also predict that the deposition efficiency has the scaling form
\begin{equation}
    \ln\eta_d=A+B\frac{1}{\delta^{1/2}}
    ~~~~ \delta \ll 1.
\end{equation}
In \Figref{fig:width_scaling} we compare the results of numerical calculations with a fit of this form. Again the agreement is excellent.

A final check is on the match between the numerics and the analytics; the analytical mathematics relies on an expansion in $x$ and $y$ and so is only valid near the stagnation point where $x,y \ll 1$. We are interested in the deposition efficiency $\eta_d(\delta)$. The deposition efficiency is set by the trajectory with the largest initial Cartesian $y_C$ coordinate, call it $y_{C0}$, at large and negative Cartesian $x_C$ (far upstream of the cylinder) for which a collision occurs. There the particle velocity is taken to be that in flow field, which is close to $U$ along the Cartesian $x_C$ axis.

\begin{figure}
  \centering
 \includegraphics[width=\linewidth]{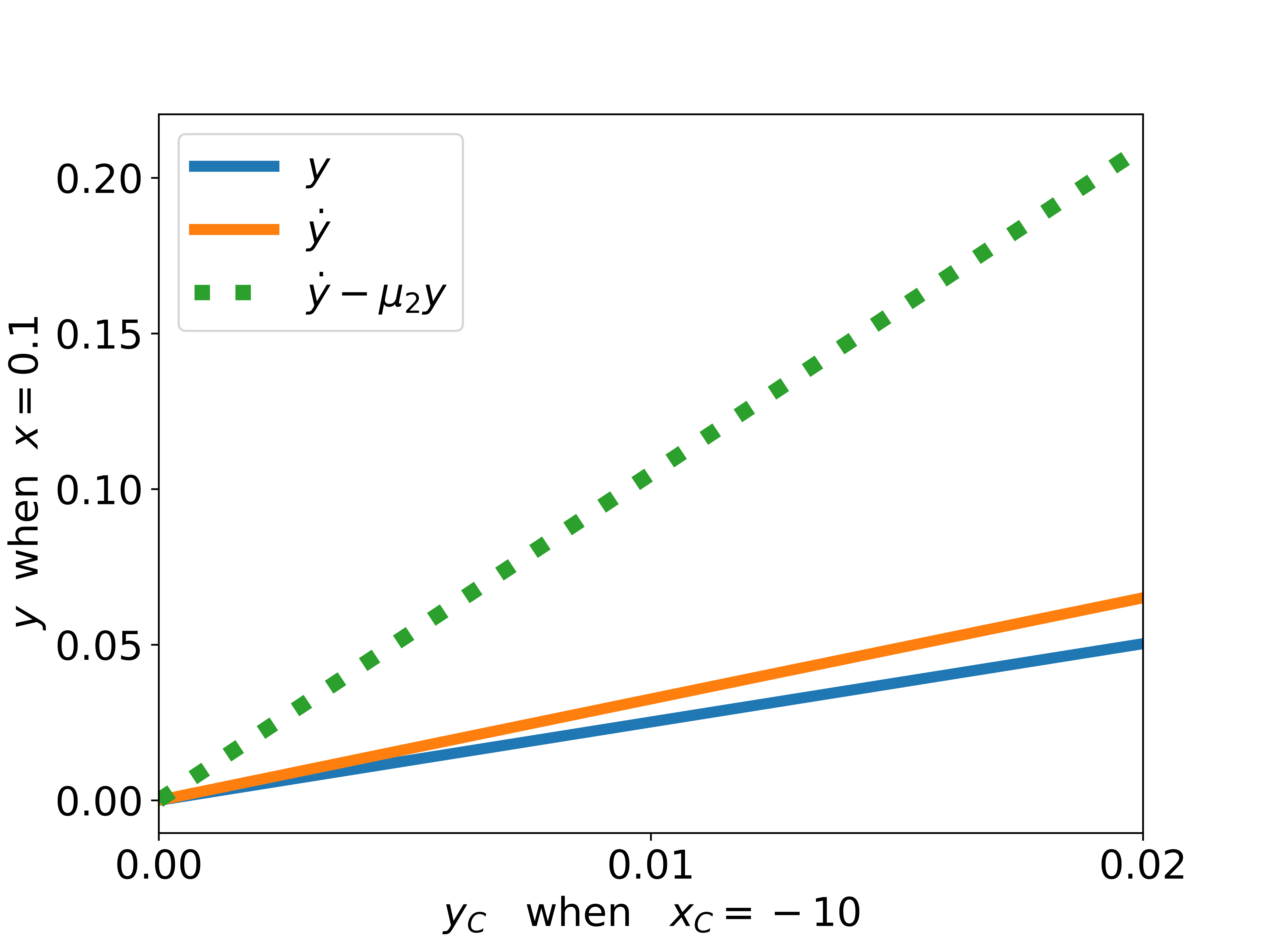}
  \caption{Plot of $y$, $\dot{y}$ and their combination $\dot{y}-\mu_2y$ which sets $C_0$, as functions of the initial off-axis deviation in Cartesian coordinates. All trajectories start at $x_C=-10$, and end at $x=0.1$. The Stokes number is $\St=1/4$. 
  }
  \label{fig:BCcheck}
\end{figure}

What the analytic mathematics uses is the initial boundary condition on $y$, as it appears as $C_0$, which has the scaling
\begin{equation}
    C_0\sim v_0-\mu_2y_0\sim \exp(\alpha)
\end{equation}
and in effect shows that $C_0\sim \exp(-1/\delta^{1/2})$. Here $y_0$ and $v_0$ are the initial conditions on $y$ and $\dot{y}$, respectively. 
To complete the link between the scaling we obtained for $C_0$ and $\eta_d$ we need to show that $C_0$ varies smoothly with the value of $y_{C0}$. We do this in \Figref{fig:BCcheck}. There we have plotted $y$, $\dot{y}$ and $\dot{y}-\mu_2y$ at $x=0.1$, as functions of $y_{C0}$. These are all obtained from numerical calculations. We see that they all vary smoothly with $y_{C0}$, so the scaling found for $C_0$ will translate $\eta_d$. In \Figref{fig:BCcheck} we selected $x=0.1$ as a value less than one but not $\ll 1$, plots with $x=0.05$ and $x=0.2$, are similar.

\section{Solution for on-axis case}
\label{sec:app1dshm}

The solution of \Eqref{eqn:xdd} is
\begin{equation}
    x(t)=A_0\exp\left(\lambda_1t\right)+B_0\exp\left(\lambda_2t\right)
\end{equation}
\begin{equation}
    \lambda_1,\lambda_2=\frac{-1\pm\sqrt{1-8\St}}{2\St},
\end{equation}
where $A_0$ and $B_0$ are fixed by the initial conditions. With $x(t=0)=x_0$ and $\dot{x}(t=0)=u_0$, $A_0$ and $B_0$ are given by
\begin{equation}
A_0=\frac{\lambda_2 x_0-u_0}{\lambda_2-\lambda_1}~~~{\rm and}~~~B_0=\frac{u_0-\lambda_1 x_0}{\lambda_2-\lambda_1}.
\end{equation}

For $\St<1/8$ both $\lambda$ are real (overdamped SHM solutions), and so the particle approaches the cylinder surface $(x=0)$ at a speed which decays as $\exp(-t)$, and hence there is only a collision in the $t\to\infty$ limit, i.e., no collision at finite time. A similar phenomena occurs in the $\St=1/8$ case. But if $\St>1/8$ then we have complex $\lambda$ (underdamped SHM solution), and the collision will occur in finite time. In this case the solution can be written as
\begin{eqnarray}
    x(t)&=&\exp\left[-t/(2\St)\right]\left(A_0\cos\left[\frac{\sqrt{8\St-1}}{2\St}t\right]\right.
    \nonumber\\
   &&~~~~~~~~~~~\left. +B_0\sin\left[\frac{\sqrt{8\St-1}}{2\St}t\right]\right),
\end{eqnarray}
or by using trigonometric identities
\begin{equation}
    x(t)=\psi\exp\left[-t/(2\St)\right]
    \cos\left[\frac{\sqrt{8\St-1}}{2\St}t+\phi\right],
\end{equation}
where $\psi$ is an amplitude and $\phi$ is a phase which are related to the constants $A_0$ and $B_0$, which are given by
\begin{equation}
A_0=x_0~~~{\rm and}~~~B_0=\frac{2\St \,u_0+x_0}{\sqrt{8\St-1}}.
\end{equation}
Thus
\begin{equation}
    \phi=\tan^{-1}\left(\frac{2\St u_0/x_0+1}{\sqrt{8\St-1}}\right).
    \label{eqn:phi_app}
\end{equation}
And rewriting using $\delta$
\begin{equation}
    x(t)=\psi \exp\left[-t/(2\St)\right]
    \cos\left[\frac{8\sqrt{2}\delta^{1/2}}{1+8\delta}t+\phi\right].
\end{equation}
The collision occurs when $x(\tCOLL)=0$ and is set by the cosine term, and so by the angular frequency, which scales as $\delta^{1/2}$. In the $\delta\to 0$ limit, $\phi$ in \Eqref{eqn:phi_app} tends to $-\pi/2$ (as $u_0<0$) and the collision time is set by $8\sqrt{2}\delta^{1/2}\tCOLL=\pi$, which leads to \Eqref{eqn:t_collide1D}.

\section{Solution of \Eqref{eq:xodeminimal}}
\label{appen:gsol}

In order to derive the scaling argument on the coupling term in \Eqref{eq:xodeminimal}, we consider only the exponentially growing term on the RHS from the solution \Eqref{eq:yexp}. Hence the differential equation becomes
\begin{equation}
    \ddot{x}+\frac{\dot{x}+2x}{\St}=C_0^2\mu_1^2\exp(2\mu_1t).
\end{equation}
The general solution to this equation consists of a complementary function which satisfies the homogeneous equation \Eqref{eqn:xdd} and a particular integral which satisfies the in-homogeneous RHS.

The general solution is then
\begin{eqnarray}
    x(t)&=&\exp\left[-t/(2\St)\right]\left(A_0\cos\left[\frac{\sqrt{8\St-1}}{2\St}t\right]\right. \nonumber\\ 
    && + \left. B_0\sin\left[\frac{\sqrt{8\St-1}}{2\St}t\right] \right)
    +E_0\exp(2\mu_1t),
    \label{eqn:gensol2}
\end{eqnarray}
where
\[
E_0=\frac{\mu_1^2C_0^2}{4\mu_1^2+\frac{2\mu_1}{\St}+\frac{2}{\St}},
\]
and
\begin{eqnarray}
A_0&=&x_0-E_0,\\
B_0&=&\frac{2\St\,u_0+x_0-E_0(1+4\St\,\mu_1)}{\sqrt{8\St-1}}.
\end{eqnarray}

\section{Inertia-driven collisions with a sphere}
\label{appen:sphere}

For a sphere in an inviscid axisymmetric flow, the flow field is\cite{acheson_book}
\begin{equation}
\frac{\vec{u}_{\rm sphere}}{U}=\left(1-\frac{R^3}{r^3}\right)\cos(\theta)\widehat{\vec{r}}
-\left(1+\frac{R^3}{2r^3}\right)\sin(\theta)\widehat{\vec{\theta}},
\end{equation}
which when expanded about the forward stagnation point with $U=R=1$ gives
\begin{eqnarray}
\vec{u}_{\rm sphere}&=&\left(-3x + 6x^2+ \mathcal{O}({\rm cubic~terms})\right)\hat{\vec{r}}
\nonumber\\
&&+\left(-\frac{3}{2}y +\frac{3}{2}xy+ \mathcal{O}({\rm cubic~terms})\right)\hat{\vec{\theta}}.
\label{eqn:ff_sphere}
\end{eqnarray}
As \Eqref{eqn:ff_sphere} is the same as the equivalent for a cylinder, \Eqref{eq:flowfieldxy}, apart from numerical prefactors, there is also a critical value of the Stokes number for a sphere. The different numerical factors shift it to the lower value of $\St_c=1/12$, which agrees with the result of Langmuir and Blodgett \cite{langmuir1946}. The change from a $-2x$ term in the flow field for a cylinder to a $-3x$ term for a sphere just shifts the critical value down by a factor of $2/3$. The equivalent of  \Eqref{eqn:xdd} for a sphere just has a $-3x$ term in place of the $-2x$ term. This means that we have the same $1/\delta^{1/2}$ scaling of the time to collide.

In spherical polar coordinates, the acceleration terms along a constant azimuthal angle trajectory, in \Eqref{eq:stokes-newton} are the same as those for the two-dimensional cylindrical polar coordinates. Hence the left-hand-sides of \Eqref{eq:stokes-newton2}, and thus of \Eqref{eq:odex_exp} and \Eqref{eqn:y}, are the same as presented here, i.e. the $\dot{y}^2$ term is the significant term, and the analysis presented for the cylinder can be repeated for the sphere. The result is that the scaling law
for the deposition efficiency has the same $\delta$ dependence as for the cylinder except with modified multiplicative constants.


%

\end{document}